\documentclass[12pt,preprintnumbers]{article}
\usepackage[utf8]{inputenc}
\usepackage{amsmath, amssymb, wasysym, slashed, multirow,hyperref}
\usepackage{pdfpages}
\usepackage{cite}
\usepackage{times}

\newenvironment{sciabstract}{%
\begin{quote} }
{\end{quote}}

\newcounter{lastnote}

\usepackage{fancyhdr}
\fancypagestyle{plain}{%
	\fancyhead[R]{RBI-ThPhys-2020-19,
ACFI-T20-06}

}

\title{Non-Wilsonian ultraviolet completion via transseries}

\author
{Alessio Maiezza,$^{1\ast}$  Juan Carlos Vasquez$^{2\dagger}$\\
\\
\normalsize{$^{1}$Ruder Bo\v skovi\'c Institute, Bijeni\v cka cesta 54, 10000, Zagreb, Croatia,}\\ \\
\normalsize{$^{2}$Amherst Center for Fundamental Interactions, Department of Physics,}\\
\normalsize{University of Massachusetts, Amherst, MA 01003, USA.}\\
\\
\small{ E-mail: amaiezza@irb.hr$^{\ast}$,jvasquezcarm@umass.edu$^{\dagger}$}
}

\date{}

\begin{document}

\baselineskip16pt %24 is the original


\maketitle


\begin{sciabstract}
We study some of the implications for the perturbative renormalization program when augmented with the Borel-Ecalle resummation. We show the emergence of a new kind of non-perturbative fixed point for the scalar $\phi^4$ model, representing an ultraviolet self-completion by transseries. We argue that this completion is purely non-Wilsonian and it  depends on one arbitrary constant stemming from the transseries  solution of the renormalization group equation. On the other hand, if no fixed points are demanded through the adjustment of this arbitrary constant, we end up with an effective theory in which  the scalar mass is quadratically-sensitive to the cut-off, even working in dimensional regularization. Complete decoupling of the scalar mass to this energy scale can be used to determine a physical prescription for the Borel-Laplace resummation of the renormalons in non-asymptotically free models. We also comment on possible orthogonal scenarios available in the literature that might play a role when no fixed points exist.
\end{sciabstract}

\section{Motivation}

Concepts of ultraviolet (UV) completion and renormalization are intimately related to each other. The latter is indeed a systematic way to keep Quantum Field Theory (QFT) finite in the UV region and, in doing this, probing also the UV fate of a given theory. One of the fundamental analytical tools to deal with renormalization is perturbation theory and, particularly important in this context is dimensional regularization upon the which renormalization program is usually implemented. A well-known feature of dimensional regularization is the insensitivity to quadratic divergences (of any energy-mass parameter). Only logarithmic divergences are present and renormalization consists of reabsorbing them in the Bogoliubov's counterterms, which in turn manifest themselves through the running of the couplings. As a result, a renormalized theory is decoupled from UV scales and one can work at some energy without concern about physics at much higher scales. Put in this way, one can state that the so-called hierarchy problem is even not defined in the usual renormalization procedure.

Things, however, may be more complicated than this standard picture. The perturbative renormalization is based on asymptotic series and thus the procedure is well-defined only for infinitesimal couplings. For finite couplings instead, these series have to be ``regularized", which technically speaking, means they have to be Borel-Laplace (BL) resummed. In principle, this enables one to extend the renormalization program to finite coupling(s). Unfortunately, this approach is hampered by the presence of the renormalons~\cite{tHooft:1977xjm}, i.e. poles on the path of Laplace integral that make ambiguous the BL resummation. While we shall go technically through this issue in the rest of the article, here we wish to emphasize the following message: the presence of renormalons obscures the idea of renormalization because they make it to depend  on the size of the coupling constant. This enhances the importance of discussing a resurgent approach for renormalization because the renormalons make fuzzy the border between a renormalizable and a non-renormalizable model. To clarify these last statements, consider for example the general non-renormalizable Lagrangian~\cite{Weinberg:1980gg} for a scalar field with $Z_2$ symmetry $\phi\rightarrow -\phi$
\begin{equation}\label{L_Weinberg}
\mathcal{L}= \frac{1}{2}\partial_\mu \partial^\mu \phi - m^2 \phi^2+ g \phi^4+ g_6 \phi^6+ g_8 \phi^8+...
\end{equation}
The inclusion of all the higher-order operators~\footnote{Mixed operators such as $\phi^3 \Box \phi, (\Box \phi)^2$, etc. can be eliminated in favor of the ones in Eq.~\ref{L_Weinberg} through a  field redefinition~\cite{Weinberg:1980gg}.} is the basis for the asymptotic-safe scenario and the search of non-perturbative fixed points~\footnote{The idea of an ultraviolet completion via fixed points was first proposed in Ref.~\cite{Wilson:1973jj}.}. The approach was proposed by Weinberg to apply it to the Einstein-Hilbert action since gravity is indeed non-renormalizable. The important point to be stressed is that a formally equivalent Lagrangian to the one in Eq.~\eqref{L_Weinberg} was proposed by Parisi to take care (perturbatively) of the renormalon ambiguities. The only difference is that the coefficients of the higher dimensional terms are exponentially suppressed for small couplings in the case of the renormalons~\cite{Parisi:1978iq}, while there is a priori no hierarchy in the ones of Eq.~\eqref{L_Weinberg}. This is sufficient to exemplify the general connection between renormalons, coupling size, and non-renormalizability.

In this work we go beyond the perturbative approach of Ref.~\cite{Parisi:1978iq} by using resurgent methods~\cite{EcalleRes:book}. We will get a non-Wilsonian modification of the standard $\phi^4$, which means that such modification cannot be obtained by integrating out heavy degrees of freedom, in contrast to what happens for Eq.~\eqref{L_Weinberg}. More in general, based on the Refs.~\cite{Maiezza:2019dht,Bersini:2019axn} and the notion of resurgence, the scope of the present article is to study the consequences for perturbative renormalization when the perturbative series are turned into transseries. Resurgence  is built upon the idea that \emph{genuine} non-perturbative information can be gotten from perturbative expansions~\cite{Ecalle1993,Costin1995} (for reviews  of resurgence in QFT see Refs.~\cite{Dorigoni:2014hea,Dunne:2016nmc,Aniceto:2018bis}).  In QFT, this is equivalent to the statement that the renormalizable Lagrangian is more fundamental than the generic non-renormalizable one. When the renormalons are unambiguously resummed~\cite{Maiezza:2019dht} via the isomorphism in Refs.~\cite{Costin1995,costin1998,CostinBook}, different possibilities open up for the UV behavior of a QFT. It is convenient to anticipate here some results: quadratic UV sensitivity for the scalar mass emerges even in dimensional regularization when perturbative renormalization is augmented with transseries; more interesting, there is the possibility to have a non-perturbative UV fixed point in the scalar model, making it ultraviolet self-complete in a non-Wilsonian sense. Different scenarios correspond to different arbitrary choices of this new parameter,  which necessarily arises in the generalized (Borel-Ecalle) resummation of the renormalons.

We should stress that in the original lattice computation, no fixed points were found for the $\phi^4(x)$ model~\cite{Wilson:1973jj} but it was argued that the inclusion of higher dimensional operators might change this conclusion. Recently, in Ref.~\cite{Jian:2020gcm} this issue was indeed verified, and a non-gaussian UV fixed point was found even at the one-loop level by adding a $\phi^6$ term. In this work,  we show that the arbitrary constant stemming from the transseries solution of the renormalization group equation can also lead to the existence of a non-gaussian UV fixed point. In this case, however, no additional higher dimensional operators are needed. The theory self-protects through its transseries structure and a Wilsonian UV completion is not necessary to avoid the Landau pole in the deep ultraviolet region.
Moreover, in the case of no fixed point, we comment on the possibility that complementary mechanisms may be at work, such a Classicalization~\cite{Dvali:2010jz}, which is a concept inspired by gravity and based on a non-Wilsonian UV completion.

The article is organized as follows. In Sec.~\ref{Resurgent approach} we discuss the resurgence approach for ordinary differential equations and its connection with QFT. In this section we refine the derivation of the non-linear differential equation behind renormalons given in Ref.~\cite{Bersini:2019axn}. This argument is fundamental because it provides the theoretical basis for the discussion in the next sections. In Sec.~\ref{scalar model} we discuss the scalar $\phi(x)^4$ model and the impact of the Borel-Ecalle resummation of the renormalons on the two-point Green function. In Sec.~\ref{prescription}, we discuss two possible applications for the merging of the perturbative renormalization with transseries. We show how the ultraviolet sensitivity of the scalar mass to a high energy scale emerges in dimensional regularization and, in particular, we discuss the case of a possible non-perturbative fixed point for the $\phi(x)^4$ model when augmented with transseries. Finally, in Sec.~\ref{UV-scenarios}, we discuss possible relations and interplay with other UV scenarios proposed in the literature.

\section{Resurgent approach}\label{Resurgent approach}

Recently, novel ideas on the renormalons have been put forward, based on the resurgence of ordinary differential equation~\cite{Maiezza:2019dht,Bersini:2019axn}. The skeleton diagrams has been resummed in a generalized sense, i.e. through a generalized Borel-Laplace summation~\cite{Costin1995,costin1998,CostinBook}, with a one-parameter transseries~\cite{Maiezza:2019dht}). The choice of single parameter transseries has been justified from the underlying equation; the renormalization group equation is of the first order. The argument has been completed in Ref.~\cite{Bersini:2019axn}, in which it has been shown that a non-linear ordinary differential equation (ODE) for the anomalous dimension $\gamma$-function can be extracted from the RGE, making robust the use of the generalized resummation of Ref.~\cite{Costin1995}.

In this section we merge and revisit the results of Refs.~\cite{Maiezza:2019dht,Bersini:2019axn} which are fundamental for the rest of the article.

\subsection{Notation on Borel-Laplace resummation}\label{basic}

Before moving on, let us define some notation. Given an asymptotic series in powers of $1/x$
\begin{equation}
s(x) = \sum_{n=0}^{\infty} a_n x^{-n}\,,
\end{equation}
we denote its standard BL resummation as
\begin{equation}\label{Laplace}
\mathcal{BL}[s(x)]=\int_{0}^{\infty} e^{-z x} B(z)  \,,
\end{equation}
being $B$ the Borel transform of $s$, $B(z)= \mathcal{B}[s(x)]$. So, rather than summing $s(x)$ one sums the more convergent $\mathcal{B}[s(x)]$ in Borel plane
\begin{equation}
B(z)= \sum_{n=0}^{\infty}  \frac{a_n}{(n-1)!}z^{n-1} \,,
\end{equation}
and then goes back through the Laplace integral in Eq.~\eqref{Laplace}. If $B(z)$ does not contain singularities in the positive real axis, i.e. along the integration path, $s(x)$ is BL resummable that is $\mathcal{BL}[s]$ is a well-defined and finite expression. In a more ``physical" language, one would say  the result is non-perturbative in the sense that it holds for any value of the parameter $1/x$, while the original asymptotic expansion $s(x)$ is valid only for $(1/x)\rightarrow 0$.

Unfortunately, series in QFT are often not BL resummable because of the presence of $n!$ behavior, which indeed brings singularities on the positive real axis of the Borel transform, i.e. along the integration path. In such a situation some generalization beyond BL is called for, and this is the goal of resurgence.

\subsection{Highlights on the resurgence and ODE}\label{basic_ODE}

The main observation is based on the expansion of the ODE~\cite{Costin1995}
\begin{equation}\label{ODE_before}
y'(x)=f(x,y(x)) \,,
\end{equation}
for small $y$ and large $x$, leading to
\begin{equation}\label{ODE_prototype}
y'(x)=f_0(x)-q y(x)+\frac{u}{x} y(x)+h(x,y(x)) \,,
\end{equation}
with $h(x,y)=\mathcal{O}(x^{-2},y^2,x^{-2}y)$. We will be interested in the case with $q>0$ and $x \in \mathbb{R}$.
It can be shown that the Borel transform $Y(z)=\mathcal{B}[y]$ is a holomorphic function in the whole complex plane except for a cut on the positive semi-axis where there are infinite singular points in
\begin{equation}\label{singularities}
S=\{z\in \mathbb{Z}, z=n q   \} \,.
\end{equation}
For later convenience, we make a change of variable converting the above expansion in $x$ to a small parameter $g = 1/x$, to identify it as a coupling constant in QFT. Eq.~\eqref{ODE_prototype} becomes then
\begin{equation}\label{ODE_prototype_g}
y'(g)=-\frac{f_0(g)}{g^2}+q \frac{y(g)}{g^2}-\frac{u}{g} y(g)+h(g,y(g))\,.
\end{equation}
Although there are infinite singularities in the set $S$, and thus on the path of the Laplace integral, the solution of Eq.~\eqref{ODE_prototype} is unambiguous module a single arbitrary constant, related to the boundary condition of ODE. This enables one to resum a $n!$-growing series with a single parameter transseries~\cite{Costin1995}, as done for the renormalons series in Ref.~\cite{Maiezza:2019dht}.

The transseries formal solution of Eq.~\eqref{ODE_prototype} is
\begin{equation}\label{transseries}
\tilde{y}(x) = \tilde{y}_0(x)+ \sum_{k=0}^{\infty}C^ke^{-(k\, q) x}x^{k\, u} \tilde{y}_k(x),
\end{equation}
which is valid in the region where $|C e^{- q x} x^{u} | < c^{-1}$, being $C$ an arbitrary constant and $c$ is given by the $|y_k(x)|\leq c^k$ for all $x$ (see Sec. 5.7 of Ref.~\cite{CostinBook}).

Heuristically, the typical result of the generalized resummation is that one captures a genuine nonperturbative piece, related to the solution of the homogeneous part of Eq.~\eqref{ODE_prototype} and thus not calculable in term of a regular series. To keep direct contact with the previous subsection, one may say that the power series expansion in $1/x$ of $y(x)$ is not BL resumable (for the non-analyticities in $S$), but it is resummable in a generalized sense once is embedded in the framework of ODE and such a generalization is a Borel-Ecalle resummation.

\subsection{Connecting QFT with the non-linear ODE framework}\label{Subsec:QFT_ODE}

Let us consider the renormalized 1PI, two-point Green function in the form
\begin{equation}\label{G21PI}
\Gamma^{(2)}_R \equiv i \left( p^2-m^2 \right) G(L,g)
\end{equation}
with $G(L,g)$ defined as~\cite{Bersini:2019axn}
\begin{equation}\label{G_true}
G(L,g) = 1-\sum_{i=1}^{\infty} \gamma_i(g) L^i + R(g) \,.
\end{equation}
 In what follows we assume that $R$ is non-perturbative, which means  its Borel transform has at least  one singularity outside the origin, but without any assumptions about the position and the type of them. Moreover,  we assume the condition of non-resonance, which means no superposition of Stokes lines (for example instantons on top of renormalons). Under this assumption, for $R$ is the only non-perturbative piece corresponding to $y(x)$ in Subec.\ref{basic_ODE}.      The  structure of the Green function in Eq.~\eqref{G_true} is justified from the renormalization group equation analysis for QED made   in Ref.~\cite{DeRafael:1974iv} -- see for instance Eq.~(2) of Ref.~\cite{Broadhurst:1992si}. In particular, the series gives the scale expansion in $L=\ln(-p^2/\mu^2)$ and $R(g)$ is a non-perturbative contribution that cannot be ruled out by a specific choice of the energy-scale $\mu$ and then $L$. The idea of Ref.~\cite{Bersini:2019axn} is to relate $R$ to the renormalons.
This follows from the renormalization group equation
\begin{equation}\label{CS}
\left[- 2\partial_L+\beta(g) \partial_g -2\gamma(g)\right]\,G(L,g)=0\,,
\end{equation}
where the anomalous dimension and the $\beta$-function are
\begin{equation}\label{beta_gamma}
 \beta(g)= \frac{d g(\mu)}{d\log(\mu)},\,\,\,\,\,\,\,\,\,\,\,\,\gamma(g)= \frac{1}{2}\frac{d\log Z}{d\log(\mu)}= \frac{1}{2}\frac{d\log G}{d\log(\mu)} \,,
\end{equation}
where $Z$ is the wave function renormalization. The last equality follows directly from Eq.~\eqref{CS} by rewriting the first two terms as a total derivative in $\mu$ (see for instance Eq.(3.4) in Ref.~\cite{tHooft:1977xjm}).
Plugging the expression~\eqref{G_true} in Eq.~\eqref{CS}, one gets
\begin{align}
&\beta(g)R'(g) +2\gamma_1(g) -2\gamma(g) -2\gamma(g)R(g) \, +\nonumber \\
&\sum_{n=1}^{\infty}\left[\beta (g) \gamma _n'(g)-2 \gamma (g)  \gamma _n(g)-2 n \gamma _{n+1}(g)-2  \gamma _{n+1}(g)\right]L^n = 0\,,\label{Ln}
\end{align}
where the prime denotes the derivative with respect to $g$. This equation can be solved as a set of recursive relations order by order in $L$, in complete analogy with the approach adopted in Ref.~\cite{Klaczynski:2013fca}. So we have the following relations:
\begin{equation}\label{Req}
 \beta(g)R'(g) +2\gamma_1(g) -2\gamma(g) -2\gamma(g)R(g)=0 \,,
 \end{equation}
 \begin{equation}\label{Reqn}
\beta (g) \gamma _n'(g)-2 \gamma (g)
  \gamma _n(g)-2 n \gamma _{n+1}(g)-2
  \gamma _{n+1}(g)=0   \,.
\end{equation}
The Eq.~\eqref{Req} can be rearranged as follows
\begin{equation}\label{Rgen}
R'(g)  = \frac{2(\gamma(g)-\gamma_1(g))}{\beta(g)} +\frac{2\, \gamma(g)}{\beta(g)}\,R       \,,
\end{equation}
In order to bring this equation in the form of Eq.~\eqref{ODE_prototype_g}, we just need to show that $\gamma(g)-\gamma_1(g)  $ is a function of $R$. To this end consider first the case $R=0$  in Eq.~\eqref{Rgen}, implying that $\gamma(g) = \gamma_1(g)$, in agreement with the results in Ref.~\cite{Klaczynski:2013fca}.  On the other hand,  if $\gamma(g) = \gamma_1(g)$ then
\begin{equation}
R'(g)  = \frac{2\, \gamma(g)}{\beta(g)}\,R \label{Rgenlimit}\,.
\end{equation}
The solution of this equation does not have singularities in its Borel transform (see Appendix~\ref{explain_eq}).   This is  in contradiction with the initial assumption about the  Borel trasnform of $R$. Hence in general~\footnote{Note that the anomalous dimension is fixed in agreement with Eq.~\eqref{beta_gamma} and it is not in general given by  the coefficient of the $L$ term in Eq.~\eqref{G_true}, which is a feature of low order perturbation theory; see for instance the discussion around Eqs.(12.49-12.51) of  Ref.~\cite{Peskin:1995ev}.}
\begin{equation}\label{gammacond}
\gamma(g)-\gamma_1(g)=M(g,R)\,,
\end{equation}
 where  $R$ is a non-analytic function and  $M(g,R)$ is such that $\lim_{R\rightarrow0}M(g,R)=0$. In what follows, we take the conservative assumption that the function $M(R,g)$ has at least an asymptotic expansion in power series of $R$ and $g$, namely~\footnote{There is also the issue of super-exponential behavior  that would require Ecalle-accelero-resummation studied in QFT in Ref.~\cite{Bellon:2018lwy}.  }
 \begin{equation}\label{MR}
M(R,g)=q\, R(g) +\frac{1}{2}(r R(g)^2+ 2sg\,R(g)) ... \,,
\end{equation}
and thus
\begin{equation}\label{gammaR}
\gamma(g)=\gamma_{pert}(g)+q'R(g) +\frac{1}{2}(r' R(g)^2+ 2s'\,g\,R(g)) ... \,,
\end{equation}
where $\gamma_{pert}(g)$ is the asymptotic expression for $\gamma(g)$.  The constants $q'$, $r'$ and $s'$ enters at $\mathcal{O}(R^2)$ in Eq.~\eqref{Rgen}, which then do not modify the position and analytic structure of $R$.   It remains now to prove that the beta function $\beta(g)$ indeed receives a correction due to the function $R$ as well. For proving this we are going to make use of Eq.~\eqref{Reqn}. Without loss of generality we may consider the beta function of the scalar $\frac{g}{4!}\phi^4(x)$ model, which is given by
\begin{equation}\label{betafunction}
\beta( g ) = 2g\, (2\gamma(g)-\gamma_{\phi^4}(g))  \,
\end{equation}
where $\gamma_{\phi^4}(g)$ is the anomalous dimension of the local operator $\phi^4(x)$. Formally, we can now plug this expression into Eq.~\eqref{Ln} to get $\gamma_{\phi^4}$ as a function of $\gamma$, $\gamma_1$ and $\gamma_n$, and it is indeed an straightforward exercise to show that $\gamma_{\phi^4} = b  g  + e R+\mathcal{O}(R^2)$, where $b,e$  are some number.  Then from Eq.~\eqref{betafunction}  we get,
\begin{equation}\label{betaR}
\beta (g) = \beta_{pert} (g)+c\, g\, R (g) +\mathcal{O}(g\,R,R^2)\,,
\end{equation}
with $c$ some constant and $\beta_{pert} (g) $ denotes the asymptotic expression of the beta function. In what follows we will assume $c\sim 1$, since its precise numerical values does not change the results we discuss in the next sections. Finally we can plug back Eqs.~\eqref{gammacond} ,~\eqref{gammaR} and \eqref{betaR} in Eq.~\eqref{Rgen} to find the non-linear equation for $R(g)$ in the normal form presented in Eq,~\eqref{ODE_prototype_g}, namely
\begin{equation}\label{ODE_R}
R'(g)= \frac{2q}{\beta_1} \frac{R(g)}{g^2} - \frac{2(\beta_2\,q -a\beta_1)}{\beta_1^2}\frac{R(g)}{g} + \mathcal{O}(g^2, g^2\,R(g),R(g)^2)\,,
\end{equation}
where $\gamma_{1\,pert}(g) = a\, g+b\, g^2 +... $ and $\beta_{pert}(g)= \beta_1g^2 +\beta_2g^3+...\, $.  The quadratic term in $R$ has been dropped out since its specific form is not important, but the presence of a non-linearity in $R(g)$ is fundamental, since it is the source of the infinite number of singularities in the Borel transform. In fact, comparing with~\eqref{ODE_prototype_g}, one see that the Borel transform  of $R$  has therefore infinite singularities on the positive axis proportional to $2q/\beta_1$ (being $\beta_1>0$ for a non-asymptotically-free model), i.e. the renormalons, and the source of this structure stems from the non-analyticity of both the anomalous dimension and the beta function. These move to all other $n-$point Green function through Swinger-Dyson equation~\footnote{See Ref.~\cite{Bellon:2014zxa} for a study of the singularities of the solutions of the Swinger-Dyson equation. Recently in Ref.~\cite{Borinsky:2020vae} and for certain QFTs, the authors wrote the Schwinger-Dyson equations as a system of non-linear ODEs, whose transseries solution displays an infinite number of singularities in the negative real axis.} and Ward identities. The Eq.~\eqref{ODE_R} provides the bridge between renormalons and the single-parameter transseries. Hence, it provides the theoretical foundation for the \emph{uniqueness} of the Borel-Ecalle resummation of the renormalon series. The Borel-Ecalle summability of the two-point function was also proved in Ref.~\cite{Clavier:2019sph} and explicitly worked out for the Wess-Zumino model.

Notice the prediction between the position of the singularities in the Borel transform and its analytic structure.
For instance for the $\phi^4(x)$ model $a=0$ and  at one loop,  we get the position of the poles in the Borel transform at $2/\beta_1$ and it is a simple exercise to show that  simple poles always occurs in the Borel transform for this particular case. The modification from simple poles is due to $\beta_2$ (two-loop beta function coefficient) and it is precisely of the form that has been known in the literature on renormalons in QFT --see for instance Eq.~(12) in Ref.~\cite{Parisi:1978iq} and Eq.~(17) in Ref.~\cite{Gardi:2001wg}.  Interestingly, only $\beta_1$ and $\beta_2$ appears in the position of the pole and the analytic structure of the Green function, which automatically makes this result scheme independent. This is a very non-trivial check, and  these two results, we believe,  are the main ones that indicate the correctness of our approach.  The results obtained  are truly non-perturbative and therefore scheme independent.

Finally, a natural step forward is to study the possibility to have a UV fixed point, once a non-perturbative estimate of the $\beta-$ function is at hand in Eq.~\eqref{betaR}.

\paragraph{Non-universality of QFT.} A discussion of Eq.~\eqref{ODE_R} and its solution in the form of Eq.~\eqref{transseries} is now necessary.
Our findings imply the introduction of an arbitrary constant $C$ in the 2-point correlation function (and then  in all the $n$-point Green functions), as a consequence of the Borel-Ecalle resummation of the renormalons. One can thus conclude that renormalization together with resurgence lead to non-universality in QFT. For universality one means the property of a system to be modeled only by a set of parameters defined in the initial Lagrangian~\cite{WILSON197475}. The introduction of $C$ put our case outside of that definition - furthermore, this constant is not unique because it has to be characterized by the model that one is describing. As also discussed in Subsec.~\ref{basic_ODE}, this results from the link between $C$ and a boundary condition for Eq.~\eqref{ODE_R}.

It would be interesting to investigate whether this non-uniqueness might be related to the Haag's theorem~\cite{Haag:1955ev}, which states that the interactive QFT cannot be unitarily mapped to the free field case. While this is beyond the scope of this work, we only notice here that the crux for both Haag's theorem and the resurgence of RGE is the interaction: within the perturbative renormalization one removes all the infinities coming from the introduction of interaction on the top of free fields QFT, and this seems to circumvent Haag's theorem (or at least makes it not manifest; see also the review~\cite{Klaczynski:2016qru}). However,  beyond perturbation theory (in fact resurgence) external information ($C$) seems to enter in the game and this might be a symptom of the problems for the interaction picture in QFT first raised in Ref.~\cite{Haag:1955ev}.  All this said, we shall see below that the constant $C$ can be at least constrained  in some cases despite the lack of a semiclassical interpretation of the renormalons~\footnote{A semiclassical interpretation of  the (IR) renormalons can be found in  Ref.~\cite{Argyres:2012vv} but in $\mathbb{R}^3\times S_1$ space-time.}

It may be helpful doing at this point a parallel with the instantons, another known source of $(n!)$ divergence of the perturbative series.
While renormalons emerge from the procedure of renormalization itself, impeding the generalization of perturbation renormalization via the Borel-Laplace resummation, the instantons are related to the classical equation of motion (see Ref.~\cite{tHooft:1977xjm}) and can be traced back from the factorial growth of all the Feynman diagrams contributing to a correlation function.  Similar to renormalons, the instantons can cause ambiguities on the positive Borel semi-axis, but these are not a problem since they can be fixed  by semiclassical methods unlike the constant $C$.

\section{Scalar model with resummed renormalons}\label{scalar model}

The entire previous section has aimed to describe the renormalons in the framework of ODEs. The function $R(g)$ obeying Eq.~\eqref{ODE_R} and then having the proper renormalons structure, has to be identified with the nonperturbative contribution to the two-point function due to the renormalons. The way to proceed requires to consider an explicit estimation of $R(g)$ coming from the Borel-Ecalle resummation of the $n-$bubbles 't Hooft 's skeleton diagram~\cite{tHooft:1977xjm} obtained in Ref.~\cite{Maiezza:2019dht} for the $\phi^4$ model. In particular, such Borel-Ecalle resummation of the renormalons has been performed with a single parameter transseries and, as discussed above, this is justified exactly from the Eq.~\eqref{ODE_R}.

In what follows we then stick to the pure $\phi^4$ model defined by the Lagrangian
\begin{equation}\label{L_bare}
\mathcal{L} = \frac{1}{2}(\partial\phi(x))^2 -\frac{1}{2}m^2 \phi(x)^2 -\frac{ g }{4!}\phi(x)^4\,
\end{equation}
and first focus on the non-perturbative propagator correction.

\subsection{Two-point function}

The two-leg skeleton diagram has been resummed in Ref.~\cite{Maiezza:2019dht}, using as an approximation the insertion of one renormalon chain of bubbles. Taking into account the corrections from the resummation of renormalons one gets
\begin{equation}
\Gamma^{(2)}_R(p)=i(p^2-m^2)\left(1+\text{analytic terms}+C\frac{e^{-\frac{2}{\beta_1 g (\mu_0^2)}}}{1+e^{-\frac{2}{\beta_1 g (\mu_0^2)}}}\right)\,,
\end{equation}
where $C$ is an arbitrary constant, $\mu_0^2=-p^2$ and
\begin{equation}\label{landaupole}
g\left(\mu _0\right)=-\frac{g(\mu )}{1-\beta _1 g(\mu ) \log \left(\frac{\mu_0}{\mu }\right)}    \,.
\end{equation}
The ``analytic terms" come from the BL resummed perturbative pieces.
The main point that we would like to convey here is that an estimate of the skeleton diagrams implies also an estimate of the non-analytic function $R$ in Eq.~\eqref{G_true}(see also App.~\ref{app1}):
\begin{equation}\label{explicit_R}
R \simeq C\frac{e^{-\frac{2}{\beta_1 g (\mu_0^2)}}}{1+e^{-\frac{2}{\beta_1 g (\mu_0^2)}}} \,.
\end{equation}
We will make use of the widely accepted interpretation in which renormalons in perturbative expansions indicate that further terms in the form of power expansion in $Q^2/ \Lambda^2$ must be included in the expressions of physical quantities (see Refs.~\cite{Altarelli:1995kz,Zakharov:1997xs,Beneke:1998ui,Shifman:2013uka} for a more detailed discussion), where $ g $ is a non-perturbative energy scale~\footnote{A possible non-perturbative generation of a mass scale was also studied in the context of resurgence and transseries in Ref.~\cite{Bellon:2016mje}.}. These power corrections are estimated from the condition that the one-loop running coupling  diverges, namely $e^{-\frac{2}{\beta_1 g (\mu_0^2\equiv Q^2)}}= Q^2/ \Lambda^2$, where take the choice of renormalization scale $\mu_0^2 =-p^2\equiv Q^2$ (see also Ref.~\cite{Shifman:2013uka}). Hence, Eq.~\eqref{landaupole} can be written as
\begin{equation}
\Gamma^{(2)}_R= i(p^2-m^2)\left(1- C \frac{p^2}{ \Lambda^2-p^2}\right)\equiv \Gamma^{(2)}_{s}+C \, \bar{ \Gamma}^{(2)}   \,,
\label{landaupoleP}
\end{equation}
where we have split in the last step the standard part and the new one proportional to $C$ and understood the radiative corrections that we are irrelevant for our discussion.

\subsection{Non-local scalar Lagrangian}

Once the resummed renormalons corrections are taken into account, the lagrangian can be rearranged as
\begin{equation}\label{L_NL}
\mathcal{L}_{NL} =\mathcal{L}_0 +\Delta\mathcal{L} =\mathcal{L}_0 +\Delta\mathcal{L}_L+ \Delta\mathcal{L}_{NL} \,,
\end{equation}
where $\mathcal{L}_0$ is the bare lagrangian defined in Eq.~\eqref{L_bare}, $\Delta\mathcal{L}_{L}$ denotes the standard \emph{local} Bogoliubov counterterm and $\Delta\mathcal{L}_{NL}$ is a new non-local counterterm.
The energy dependence in Eq.~\eqref{landaupoleP} can be indeed traded in a non-local kinetic operator via Fourier transform
\begin{equation}\label{L_CT}
\Delta\mathcal{L}_{NL} = C \, \int d^4y \int d^4 p e^{-i p (x-y)} \bar{ \Gamma}^{(2)}(p)  \equiv \int d^4y \phi(x) F(x-y) \phi(y)   \,.
\end{equation}
The non-local piece coming from the non-perturbative correction can be explicitly worked out in the \emph{static} frame as
\begin{align}\label{L_CT-static}
\Delta\mathcal{L}(x)_{NL}= & \frac{1}{2}C   \Lambda^2 \left( \partial_{\mu}\phi(t,\vec{x})\int d^3\vec{y}\,\,\, \frac{e^{- g \, |\vec{x}-\vec{y}  |}}{8\pi^2 |\vec{x}-\vec{y}  |} \partial^{\mu} \phi(t,\vec{y})\right. \, \\ \nonumber
&\left. - m^2  \int d^3\vec{y}\,\,\, \frac{e^{- g \, |\vec{x}-\vec{y}  |}}{8\pi^2 |\vec{x}-\vec{y}  |} \phi(t,\vec{x})\phi(t,\vec{y}) \right)\,.
\end{align}
This new piece must be interpreted as the type of \emph{counterterm} proposed in Ref.~\cite{tHooft:2002pmx} to describe what is called perturbative confinement. The idea of Ref.~\cite{tHooft:2002pmx} is to start with an unusual counterterm as an ansatz to take into account some non-perturbative quantum effects, coming from some unspecified higher-order corrections (modeling the confinement in that case) and, once such counterterm has been considered, one can proceed perturbatively ``returning" the new piece order by order. The reason is that, when doing loop computations, one uses $\mathcal{L}_0+\Delta\mathcal{L}$ as the lowest order approximation. In practice, this means that one must include in the loops the modified propagator including the new correction coming from the non-local counterterm. This is precisely what happens here in Eq.~\eqref{L_CT}, but, rather than an ansatz, in our case, this emerges through the resurgence when calculating the non-perturbative effects from the renormalons. In the same spirit of Ref.~\cite{tHooft:2002pmx}, we shall study the consequences of this approach in the next section.

\section{Scale invariance and other possible prescriptions for the single parameter transseries} \label{prescription}

The scope of this section is to constrain the otherwise free transseries parameter $C$ of the Borel-Ecalle renormalons resummation. The main result that we shall show is scale invariance at high energy for proper values taken by the constant $C$.

In general, there are two parallel scenarios:
\begin{itemize}

\item in the first scenario, the effect of the resummed renormalons is driven by dimensional transmutation as discussed in the previous section, i.e. $e^{-\frac{2}{\beta_1 g (Q^2)}}= Q^2/ \Lambda^2$ and the theory is defined only up to $\Lambda$. In this case, we evaluate the one-loop scalar mass correction, showing a high scale sensitivity. If one demands the decoupling limit on such a mass correction, this provides a trivial solution $C=0$. The case with $C\neq 0$ and no UV fixed point is worthy of a separate discussion that we shall give in section~\ref{UV-scenarios};

 \item in the second scenario, we find a non-perturbative UV fixed point. In this case, $ \Lambda \rightarrow \infty$, thus there is not anymore dimensional transmutation nor a hierarchy problem as in the previous case. Needless to say, the presence of a UV fixed point is by far the most interesting case, since the model becomes ultraviolet self-complete in a non-perturbative sense. Such ultraviolet self-completeness is a central point of this work. However, the requirement scale invariance does not still fix $C$ uniquely but restricts the range of possible values that it may take.

\end{itemize}

These two points shall be analyzed in Subsecs.~\ref{mass_correct_decoupling}, ~\ref{UV_fixed_point_case} respectively.

\subsection{Mass correction, modified propagator and decoupling limit}\label{mass_correct_decoupling}

From Eq.~\eqref{landaupoleP}, the propagator is of the form:
\begin{equation}
G(p) = \frac{i}{(p^2-m^2)\left(1-C \frac{p^2}{\Lambda^2-p^2}\right)} = A_1 \frac{i}{p^2-m^2}-A_2 \frac{i}{p^2-\frac{\Lambda^2}{C+1}} \,
\end{equation}
where
\begin{equation}
 A_1 = \frac{\Lambda^2-m^2}{\Lambda^2-(C+1) m^2}\,,  \,\,\,\,\,\,   A_2 = \frac{C  \Lambda^2}{(C+1)[\Lambda^2-(C+1) m^2] }\,.
\end{equation}
The result is thus the sum of the standard propagator plus another propagator with a modified mass square $\Lambda^2/(1+C)$. The propagator resembles the Pauli-Villars regulator, but unlike that case, it does not cancel the quadratic divergences in the scalar mass for any finite $C$. Using the modified propagator, the one-loop correction to the scalar mass at scale $\mu=m$ and after a $\overline{MS}$ subtraction is
\begin{equation}
(m^{1-loop}_{finite})^2= \frac{g \left[A_1 (C+1) m^2+A_2 \Lambda^2 \log \left(\frac{(C+1) m^2}{ \Lambda^2}\right)+A_2  \Lambda^2\right]}{32 \pi^2 (C+1)} \,.
\end{equation}
We can explicitly see a finite correction proportional to $\Lambda^2$. This is in contrast to the common lore that no quadratic mass corrections arise in dimensional regularization. This is a \emph{genuine} non-perturbative effect coming from the Borel-Ecalle resummation procedure.

One would be tempted to remove this sensitivity by going to another renormalization scheme in which for example the entire correction is reabsorbed in the counterterm $\delta m$, but this is not a loophole because in such a case the quadratic correction enters into the beta function of $m$
\begin{equation}
\beta_{m^2}= \mu\frac{d m^2}{d\mu} \supset -\frac{C g }{16\pi^2 (C+1)^2}  \Lambda^2 \,,
\end{equation}
which immediately brings back the $ g ^2$ correction to $m^2$.

Therefore, when $C\neq 0$, there would be corrections proportional to $ g ^2$. This is nothing but the hierarchy problem that has been brought
into dimensional regularization from the generalized resummation together with dimensional transmutation. Fortunately, this formalism also offers a \emph{technical} solution because the new quadratic piece is proportional to the arbitrary constant $C$.

Therefore, a \emph{technical} way-out is that one requires the decoupling of the heavy scale from the physical scalar mass, imposing the condition $C=0$. This is a possible prescription for the Borel-Laplace summation of the UV renormalons, trivially consisting of taking the Cauchy principal value of the Laplace integral. With this condition, one is in the completely standard case: no hierarchy problem and the usual Wilsonian UV completion is required above the Landau pole energy scale. However, one must keep in mind that this is not the only possibility. Less standard alternatives are possible with $C\neq 0$, related to non-Wilsonian UV completions of the non-local model discussed above. We shall further comment on this issue in Sec.~\ref{UV-scenarios}, while in the next subsection we present our specific proposal for a non-Wilsonian UV completion, based on non-perturbative UV fixed points captured by Borel-Ecalle resummation, and with an absence of a cutoff. Not less important, we shall stress on the reason why we regard as non-Wilsonian this kind of UV completion from transseries.

%%%%%%

\subsection{Non-perturbative fixed point in the scalar model}\label{UV_fixed_point_case}

As pointed out in Ref.~\cite{Parisi1977}, there are at least three instances in which fixed points exist. The first one is when the value of the coupling constant at the critical value $ g _c$ is very small, i.e.  $ g _c \sim \epsilon$ in $4-\epsilon$ space-time dimensions with $\epsilon\ll1$~\cite{Wilson:1973jj}. There is a second option in which $ g _c\sim 1$ and $\epsilon\sim 1$, such as the  $\phi^4$ model in three-dimension. In this case, the fixed point already exists at one loop and higher-order corrections in both loops and $\epsilon$ improves the value of $ g _c$. There is a third case in exactly four dimensions in which the beta function cannot have a zero at the one-loop level and thus all the higher-order corrections must be estimated. In doing that,  one has to resort to approximants, such as Borel-Pad\'e or the hypergeometric Meijer G-function~\cite{Mera:2018qte}, as done in Ref.~\cite{Antipin:2018asc}. There is, however, a fourth option in which a fixed point may be found by canceling the 1-loop beta function with a flat contribution~\cite{Klaczynski:2013fca}. As already stressed many times, ``all orders in perturbation theory" is not a well-defined expression when one is dealing with non-BL resummable series. Once the $(n!)$-order divergences due to renormalons are properly taken into account in the analyzable function framework~\cite{CostinBook}, the transseries parameter $C$ can be used to find new zeros for the beta function (for example in the $\phi^4$ model).
In analogy with the epsilon expansion, $C$ takes the same role as the $\epsilon$ parameter in  $4-\epsilon$ dimensions.

As anticipated, we show that the model may have a non-gaussian fixed points in four space-time dimensions. Let us start with Eq.~\eqref{betaR}
\begin{equation}\label{b_eff}
\beta_{eff}=\beta_{pert}+ g  R= \beta_1  g ^2+  g  R+\mathcal{O}( g ^2|R^2)  \,,
\end{equation}
and now use the fact that $R$ has been explicitly estimated via the renormalon resummation in Eq.~\eqref{explicit_R}. A nontrivial fixed point can be found by requiring
\begin{equation}\label{criticality}
\beta_{eff}=0 \,\,\,\,\,\,\,\,\,\, \Rightarrow C( g _c)=-\beta _1  g  _c \left(e^{\frac{2}{\beta _1  g _c}}+1\right) \,,
\end{equation}
being $ g _c$ the value of the coupling at the critical point. It is easy to realize  from Eq.~\eqref{criticality} that $C( g _c)$ is smaller than zero for the fixed point to exist   and  that it has also a maximum  allowed value  which corresponds to a lower limit on $|C( g _c)|$ and hence
\begin{equation}\label{Cmin}
|C( g _c)|\geq\frac{2}{W\left(e^{-1}\right)}  \,,
\end{equation}
And the lower limit is reached for the following value of the coupling
\begin{equation}\label{lambda_c_Cmax}
( g _c)_{max}^{UV}= \frac{2}{\beta _1 \left(W\left(e^{-1}\right)+1\right)} \,,
\end{equation}
where  $W$ is the Lambert function and $ g _c <( g _c)_{max}^{UV} $ for all the UV fixed points.

One now has to make sure that the transseries solution is consistent for the value of the coupling $g _c$ considered. The Eq.~\eqref{criticality} gives a reliable solution for the non-linear ODE expansion in the region where the condition~\cite{CostinBook} (see subsection~\ref{basic_ODE} below Eq.~\eqref{transseries})
\begin{equation}\label{bound}
\left| C ( g _c)\right| <2 e^{\frac{2}{\beta _1  g  _c}}  \,
\end{equation}
is satisfied. Using Eq.~\eqref{criticality} we then find the condition
\begin{equation}
\beta _1  g  _c \left(e^{\frac{2}{\beta _1  g _c}}+1\right)- 2 e^{\frac{2}{\beta _1  g  _c}}<0 \,,
\end{equation}
and then
\begin{equation}
 g _c^*   <   \frac{2}{\beta_1 (1+W(e^{-1}))}    \approx  82.35     \,,
\end{equation}
thus $( g _c)_{max}^{UV}$ in Eq.~\eqref{lambda_c_Cmax} has to be discarded.

So far, we have used the one-loop approximation for the perturbative beta-function to make our point as clear as possible. However, one could consider a BL resummed $\beta_{pert}$ as well. In the next paragraph, after this improvement, we shall reassess $ g _c\leq ( g _c)_{max}^{UV} $ and it shall be consistent with the bound in Eq.\eqref{bound}.

\begin{figure}
 \centering
 \includegraphics[width=1\columnwidth]{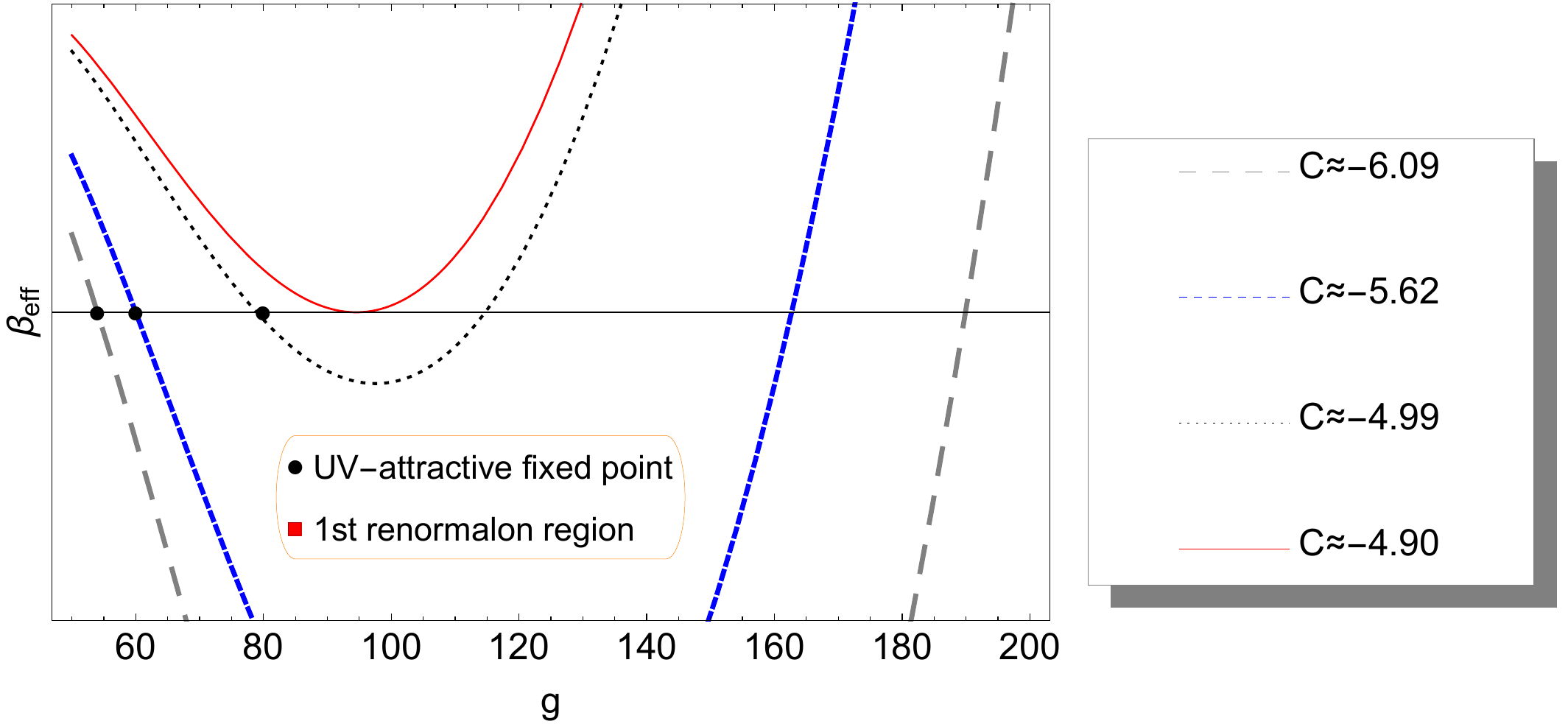}
 \caption{Non-perturbative beta function  $\beta_{eff}$ as a function of the critical coupling $ g _c$ . The color lines represent different values of the constant $C$ consistent with Eq.~\eqref{range}. The solid red line corresponds to the maximum $C$. The bullets denote the UV attractive fixed points and the zero of the solid red line needs a separate discussion (see text). The red square is the region below the first renormalon: the nonanalytical term $\exp(-2/(\beta_1  g ))$ is progressively suppressed inside that zone.}
 \label{beta_function}
\end{figure}

\paragraph{Improvement from BL resummation of the perturbative part.}
In the resurgent approach,  one starts with the zero-order  term (in $C$) of the transseries in Eq.~\eqref{transseries}, which is the principal value of the standard Borel-Laplace result ($\mathcal{BL}[\beta_{pert}]$). Then, the piece $\propto R$ in Eq.~\eqref{b_eff} provides the topologically disconnected contribution from the origin  of the Borel plane ($z=0$),  which is the only non-zero one for  simple poles the Borel transform. The $\mathcal{BL}[\beta_{pert}]$ piece,  being connected to the origin, can be  then estimated  using the state-of-the-art truncated loop expression augmented with some approximation method.

Technically, this has to be done algorithmically, as through the Pad\'e approximants or the recent and fast-convergent hypergeometric Meijer-G function approximants~\cite{Mera:2018qte}. Ideally, one would have to use the effective charge beta function~\cite{PhysRevD.29.2315,BENEKE1993154}, but building an approximated $\mathcal{BL}[\beta_{pert}]$ in a given scheme is sufficient to prove the presence of fixed points, which are scheme-independent.

For our scope, it is sufficient to employ the state-of-the-art resummed result for $\phi^4$ model provided in Ref.~\cite{Antipin:2018asc} and based on $\overline{MS}$, so that
\begin{equation}
\beta_{pert} \mapsto \beta^{MG} \,,
\end{equation}
where the index $``MG"$ stays for Meijer-G function and the expression for $\beta^{MG}$ becomes semi-numerical
\begin{equation}\label{MG_beta}
\beta^{MG}=  g  ^2 \left(0.019-1.2\times 10^{-19}  g  G_{3,4}^{4,1}\left(\frac{189.133}{ g  }|
\begin{array}{c}
 1,2.99191,0.0577911 \\
 1,1,18.8477,0.0631126 \\
\end{array}
\right)\right)       \,.
\end{equation}
With this improved of $\beta_{pert}$ to be used in Eq.~\eqref{b_eff}, we get the critical values
\begin{equation}\label{final_critical}
C^{max}\thickapprox -4.90 \,\,\,\,\,\,\,\,\,\,\, ( g _c)_{max}^{UV}\thickapprox 94.53 \,.
\end{equation}
The condition of criticality give the following constrain for  the allowed values of $|C|$
\begin{equation}\label{range}
|C( g _c)| \geq 4.90   \,,
\end{equation}
and
\begin{equation}
 g _c  <  g _c^*   \approx  115  \,.
\end{equation}
Therefore, after the MG-improvement of the beta function, we have found that the maximum critical value in Eq.~\eqref{final_critical} satisfies Eq.~\eqref{bound}.

The requirement of finiteness, i.e. the absence of a Landau pole, has required a restriction to the otherwise free parameter $C$ emerging from the generalized resummation. The parameter, however, does not remain uniquely fixed. There is a whole range of possibilities for example in Fig.~\ref{beta_function}, which shows the beta function for some values of the constant $C$ consistent with Eq.~\eqref{range}.

In summary, one sees that there is a range of UV attractive fixed points depending on the value of the constant $C$. As a benchmark, it is also shown in Fig.~\ref{beta_function} the first renormalon at $z_{1st}=2/\beta_1$, meaning that for $ g <<z_{1st}$ the nonanalytical term $\exp(-2/(\beta_1  g ))<<1$. The strength of the coupling $ g $ can be then normalized to $2/\beta_1 \approx 105$, so its absolute size has to be understood with respect to this value. There is also a range of IR attractive fixed points and the black dotted line is the border-line for these IR fixed points to satisfy the bound in  Eq.~\eqref{bound}.
Finally, notice that the extremal value $C\simeq -4.9$ is an interesting situation in which there is a UV fixed point if the physical coupling $ g  \leq  g _c$. Whereas, if $ g  \geq  g _c$ the fixed point is not reached in the far UV. Therefore, it does not fall off into the usual notion that UV fixed points can be thought of as sinks of the RG flow.

\section{UV-completeness, non-renormalizable Lagrangians and other scenarios}\label{UV-scenarios}

We have argued so far that the resurgence of the renormalization group equation may describe a non-perturbative UV self-completion and that such completion is non-Wilsonian.
Let us go through this issue in detail by starting again from Eq.~\eqref{L_Weinberg}, which represents a prototype of non-perturbative Wilsonian
UV completion. It is  well known that Eq.~\eqref{L_Weinberg} is equivalent to a renormalizable Lagrangian with two fields $\phi,\Phi$ (same logic if one considers  more fields)
\begin{equation}\label{L_twofields}
\mathcal{L} = \frac{1}{2}(\partial\phi(x))^2 + \frac{1}{2}(\partial\Phi(x))^2 -\frac{1}{2}m^2 \phi^2 -\frac{1}{2}M^2 \Phi(x)^2 -\frac{ g _1}{4!}\phi(x)^4-\frac{ g _2}{4!}\Phi(x)^4  - \alpha \phi^2 \Phi^2 \,.
\end{equation}
in the limit $M>>m,q$ being $q$ the momentum exchange in the considered processes. In this limit, the equivalence between Eq.~\eqref{L_Weinberg} and Eq.~\eqref{L_twofields} comes by integrating out the heavy field $\Phi$ and then take $M$ as a cut-off $\Lambda$ such that the $g_{2n}$ ($n\geq3$) in Eq.~\eqref{L_Weinberg} are
\begin{equation}\label{Wilson_mathc}
g_{2n}\propto \frac{1}{ \Lambda^{2 n - 4}}\,.
\end{equation}
In the following subsection, we compare this logic with the results coming from the resurgence of the renormalons.

\subsection{Resurgence and the Operator Product Expansion} We have seen above that resuming the renormalons leads (in the case of no fixed points) to a non-local counterterm that must be added into the renormalized Lagrangian. It is worth recalling that in the usual perturbation theory it is impossible to obtained non-local terms since the Bogoliubov counter-terms suffice to prove the renormalizability at any finite order. Thus at any finite order in perturbation theory, the $n!$ behavior of the perturbative expansion does not lead to divergences - the problem instead arises when $n\rightarrow\infty$ and the transseries enter into the game to cure this $n!$ behavior.

Let us focus then one the non-local piece in Eq.~\eqref{L_CT}
\begin{equation}\label{DL}
\Delta \mathcal{L}= \phi(x) \int d^4 y F(x-y) \phi(y)\,,
\end{equation}
and applying the OPE~\cite{PhysRev.179.1499}:
\begin{equation}
\phi(x)\phi(y) \sim \sum_{n=0}^{\infty} C_{2n}(x-y)\phi^{2n}(x)\,,
\end{equation}
gives
\begin{equation}\label{connect_classicalization}
\Delta \mathcal{L} \sim \sum_{2n} \phi^{2n}(x)\int d^4 y\, F(x-y)\, C_{2n} (x-y) \equiv \sum_{2n}g_{2 n}\,\phi^{2n}(x)  \,.
\end{equation}
This expression gives the non-renormalizable lagrangian in Eq.~\eqref{L_Weinberg}. However, the relation in Eq.~\eqref{connect_classicalization} is only asymptotic and in this sense OPE captures only a part of the non-local term in Eq.~\eqref{DL} which is \textit{per se} a non-Wilsonian modification of the lagrangian.

It is important to stress once again that Eq.~\eqref{L_NL} is valid only up to $ g $, and then it is not UV complete. Conversely, when there is a fixed point dimensional transmutation does not take place (see Subsec.~\ref{UV_fixed_point_case}). In this case, the coupling becomes a function of the transseries parameter $C$ and in practice, the interactive model is modified via the effective coupling
\begin{equation}\label{new_interactive_model}
 g (\mu) \mapsto  g _{eff}(\mu,C)\,.
\end{equation}
Therefore, for given values of $C$ found in Subsec.~\ref{UV_fixed_point_case} the model remains fundamental at any scale and is truly UV-complete. Note that Eq.~\eqref{new_interactive_model} follows directly from the effective beta function in Eqs.~\eqref{betaR},~\eqref{b_eff} and changes drastically the picture of the usual perturbative renormalizable $\phi^4$ model. Also the self-complete model defined from the interaction in Eq.~\eqref{new_interactive_model} cannot be interpreted in Wilsonian sense but rather in terms of transseries.

\subsection{Comments on different UV scenarios}

Stressing the notion of Wilsonian vs non-Wilsonian UV completion, in this part we point out the difference between the asymptotic safe scenario of Eq.~\eqref{L_Weinberg} and the fixed point model through resurgence. To this end, suppose one builds Eq.~\eqref{L_Weinberg} by integrating out some heavy degrees of freedom (as in Eqs.~\eqref{L_twofields} and~\eqref{Wilson_mathc}): in this case, there is a cutoff $\Lambda =M$ and the meaning of non-perturbative fixed points is not transparent. The reason  is that the notion of scale invariance is by definition in contradiction with any finite energy scale -- in this case the cut-off $\Lambda $.

This issue can be circumvented by interpreting Eq.~\eqref{L_Weinberg} with no reference to any heavy energy scale, but in this way the hierarchy of the higher-order operators provided by Eq.~\eqref{Wilson_mathc} is lost. In practice, one does not have a rationale to stop the expansion in Eq.~\eqref{L_Weinberg} and hence ``all" the operators should be equally considered. Therefore, even finding a UV fixed point from a given number of higher-order operators, one cannot guarantee that the result is not invalidated by the inclusion of additional terms. Of course, one can test in principle the stability of the result by adding just the first few terms, but the problem is never self-contained because of the lack of the hierarchy between the couplings $g_{2n}$.

Resurgence, being constructed in a mathematically robust way may provide the rationale that is lacking within the effective approach to scale invariance. In particular, sticking to the subject of the present paper, we have merged the concept of renormalization with the concept of resurgence, getting as a result a possible non-perturbative UV completion (in the subsection~\ref{UV_fixed_point_case}). Notice that the UV fixed point is built using a double expansion in $g$ and $R$ (e.g. Eq.~\eqref{b_eff}), which is formally justified by the ODE in Eq.~\eqref{ODE_before} upon which the resurgence ideas are developed.

\paragraph{An orthogonal scenario.} We have considered two separate cases in subsections~\ref{UV_fixed_point_case} and~\ref{mass_correct_decoupling}: one with UV scale invariance and another case without it. In the latter case, one ends up with a non-local scalar model, in which the non-locality is manifest at a typical  energy scale $\Lambda$. This model is not UV complete, because it is valid only up to a cutoff $\Lambda$. Nevertheless, it represents a non-Wilsonian UV modification of the standard $\phi^4$ model, which is a consequence of the incompleteness of the perturbative renormalization.
With this in mind, one may speculate whether an orthogonal mechanism such as Classicalization~\cite{Dvali:2010jz} can be invoked. Similar to the asymptotic safety paradigm, the Classicalization hypothesis is gravity-inspired but this is perhaps the only feature shared with it. The basic idea is that strong coupling is prevented by collective excitations: high momentum exchange in scattering is re-distributed in many quanta so that,  in practice, the coupling stays always in the weak coupling regime.

In the setup discussed in subsection~\ref{mass_correct_decoupling} with a cutoff $\Lambda$, one may consider two scenarios: one when $\Lambda\gg m$ and the other when $\Lambda \gtrsim m$. In the first scenario, one has a dramatic hierarchy problem that can be solved by requiring $C=0$, as already discussed in subsection~\ref{mass_correct_decoupling}. In the case  $\Lambda\gtrsim m$  the interaction becomes strong just above the energy scale $m$, turning the original $\phi^4$ model  into a non-local one. Notice that since $ \Lambda\sim m$ the constant $C$ may be different from zero. In this case, there is indeed no hierarchy problem and the only concern is how to avoid the divergence of the interaction coupling around the energy scale $\Lambda$ (see also Ref.~\cite{Dvali:2016ovn}). In this scenario, Classicalization may offer the possibility that the (non-local) model protects itself as for example in Ref.~\cite{Dvali:2010jz}, in which the standard Higgs boson is proposed as classicalizer.
We should remark that we are not studying the implementation of the Classicalization since this is out of the scope of the present work, but we are rather providing an example in which the hierarchy problem is cured even though $C\neq 0$,  unlike in subsection~\ref{mass_correct_decoupling}.

A final comment is in order. Classicalization solves the problem of strong coupling in a statistical-mechanic way, i.e. via a many-body redistribution of energy such that the scattering $2\rightarrow 2$ is suppressed (with respect than $2\rightarrow N$ with large $N$) but not impossible: strong coupling is not avoided in the strict sense. Unlike Classicalization, from our point of view centered on the renormalons, the strong coupling is the dramatic manifestation of the incompleteness of the renormalization and the notion of renormalizability (see also Ref.~\cite{Crutchfield:1979rt}). Moreover, being  conceptual, let us emphasize that this incompleteness does not distinguish between weak and strong regimes since one moves smoothly from one limit to the other.

\section{Epilogue}

We have provided a bridge between renormalizable and non-renormalizable models through the notion of renormalons, resurgence, and non-linear ordinary differential equations. Starting from the framework defined in Refs.~\cite{Maiezza:2019dht,Bersini:2019axn}, in this article, we have analyzed some relevant implications for QFT taking as a prototype the scalar $\phi^4$  model.

The main idea is that the perturbative renormalization is not complete in the sense that it is based on asymptotic and non-BL resumable series. A more general isomorphism needs to be used to get consistent results and in particular to resum the renormalons. Such an isomorphism is the Borel-Ecalle resummation that we have implemented within the framework of ODE which, thanks to the RGE, enables us to trade in a single parameter ($C$) transseries the effects of the renormalons. Through the notion of \emph{resurgence},
we have proposed how an improvement of the perturbative renormalization procedure might look like. As a result, two mutually exclusive scenarios open up.

The first one is characterized by a cutoff $\Lambda$. The interesting thing is that this scale enters via the resurgence formalism together with dimensional transmutation in the propagator and thus in loop corrections to the scalar mass. In other words, we have shown how the hierarchy problem can be formalized in dimensional regularization, which in general is known to be insensitive to the quadratic divergences. The UV modification of the standard $\phi^4$ model  is non-local and the non-locality-energy-scale is $\sim\Lambda$. We have argued that this kind of modification is non-Wilsonian. When one sets  $C=0$,  both the non-locality and the hierarchy problem go away and the Cauchy principal value prescription for the Laplace integral of the renormalons remains. In this case, however, the Landau pole is still an issue and  we have commented on the Classicalization as one interesting framework to address it. Specifically, Classicalization might be on top of the resurgence modification that we have introduced, and it may be suggested by the non-Wilsonian nature of the standard $\phi^4$ model when augmented with transseries.

The second scenario is our main result, in which we have shown the existence of UV-attractive fixed points, depending on the values of the transseries parameter $C$. In this case, there is no cutoff and the model remains consistent at any energy, therefore it is self-complete. We have argued that this completion is genuinely non-Wilsonian (and non-universal) since the behavior of the interaction $ g (\mu,C)$ is indeed drastically affected by an external parameter, $C$, emerging from the Borel-Ecalle resummation. It is worth stressing that $C$ is not uniquely determined, but rather the range of its possible values gives rise to an entire family of models that are scale-invariant in the ultraviolet region.

%%%%%%%%%%%%%%%%%%%%%%%%%%%%%%%%%%%%%%%%%%%%%%%%%%%%%%%%%%%%%%%%%%%%%%%%%%%%%%%%%%%%%%%%%%%%%%%%%%%%%%%%%%%%%%%%%%%%%%%%%%%%%%%%%%%%%%%%%%%%%%%%%%%%%%%%%%%%%%%%%%%%%%%%%%%%%%%%%%

\section*{Acknowledgement}

We thanks Arpad Lukacs for pointing us out a typo.
AM was partially supported by the Croatian Science Foundation project number 4418. JCV was supported in part under the U.S. Department of Energy contract DE-SC0015376.

%%%%%%%%%%%%%%%%%%%%%%%%%%%%%%%%%%%%%%%%%%%%%%%%%%%%%%%%%%%%%%%%%%%%%%%%%%%%%%%%%%%%%%%%%%%%%%%%%%%%%%%%%%%%%%%%%%%%%%%%%%%%%%%%%%%%%%%%%%%%%%%%%%%%%%%%%%%%%%%%%%%%%%%%%%%%%%%%%%%%

\appendix
\section{Further details on the ODE from RGE}\label{explain_eq}

We explicitly show here that the Borel transform of solution of equation
\begin{equation}\label{bis}
R'(g)  = \frac{2\, \gamma(g)}{\beta(g)}\,R   \,.
\end{equation}
does not have poles in $z>0$.

If $\beta,\gamma$ are independent from $R$, the statement is trivial so let us consider them in the form $\beta=\beta_{pert}+ \eta R +\mathcal{O}(R^2) $ and $\gamma=\gamma_{pert}+ q' R +\mathcal{O}(R^2) $. Plugging these expression in Eq.~\eqref{bis} and expanding for small $g,R$ once again according to the logic of the subsec.~\ref{basic_ODE}, yields
\begin{equation}
R'=\frac{2 R \left(a-\beta _1\right)}{\beta _1 g} +\mathcal{O}(R^2)     \,.
\end{equation}
where we have also used the perturbative expansion of $\beta_{pert},\gamma_{pert}$ given subecec.~\ref{Subsec:QFT_ODE}. Comparing with Eq.\eqref{ODE_prototype_g}, one sees that the term proportional to $R/g^2$, which determines the position of the poles, is absent. We conclude that the solution of Eq.~\eqref{bis} is analytic in the variable $g$ except at  the origin.

\section{Pocket summary on resurgent functions from ODEs}\label{app1}

For completeness, we give a summary on the generalized resummation for analyzable functions, extracted from Refs.~\cite{Costin1995,costin1998,CostinBook} and recently applied to renormalons in QFT in Ref.~\cite{Maiezza:2019dht}. The main text can be read independently from this appendix. Starting with the transseries in Eq.~\eqref{transseries} and considering the function $Y_k(z)=\mathcal{B}[y_k(x)]$; $Y_0(z)$ is then the Borel transform of the perturbative series.
Next, suppose that $Y_0(z)$ is known at all orders in perturbation theory (in true QFT this is not true of course, nevertheless, as discussed in the text, $Y_0(z)$ can be built from loop expansion together with approximants, as Pad\'e or hypergeometric ones).

\paragraph{Resurgence.} Here we borrow and condense informations as better we can from chapter 5 of Ref.~\cite{CostinBook}. Given $Y_0$, resurgence is the mechanism to reconstruct the entire function $y(x)$ in Eq.~\eqref{transseries}, through the recursive calculations of all the $Y_k$. The procedure is as follows. First define $Y^{\pm}_k(z) \equiv Y_k(z\pm i \epsilon)$, then
\begin{equation}\label{eq:resurgence}
S_0^k Y_k=(Y_0^- - Y_0^{-k-1+}) \circ \tau_k, \quad \tau_k: z \mapsto z +k\, q\,,
\end{equation}
where $S_0$ is the non-perturbative Stokes constant and
\begin{equation}\label{eq:analyticcontinuation}
Y_k^{-m+} = Y_{k}^{+} + \sum_{j=1}^{m} \binom{k+j}{k} S_{0}^{j} Y^{+}_{k+j}\circ \tau_{-j}\,.
\end{equation}
One arrives at the \emph{balanced average} associated with each $Y_k$
\begin{equation}\label{balanced:average}
Y_k^{bal} \equiv Y_k^+ + \sum_{n=1}^{\infty}2^{-n} (Y_k^- - Y_k^{-n-1+})\,.
\end{equation}
Finally, denoting the Laplace transform $\mathcal{L}(Y_k^{bal})\equiv\mathcal{E}(y_k)$, one has the neat result as
\begin{equation}
\mathcal{\sigma}(y_0(x)) \mapsto \mathcal{E}(y_0)(x) + \sum_{k=1}^{\infty} e^{-k\cdot q /x} \mathcal{E}(y_k) (x)\,,
\end{equation}
where $\sigma$ denotes the generalized operation of the Borel resummation. When no poles are present in the positive real axis the usual Borel procedure is recovered.

\paragraph{Renormalons.} It turns out from direct estimations on the skeleton diagram for $\phi^4$ that the renormalons are simple poles in the Borel positive axis~\cite{tHooft:1977xjm}. Barring for a moment the non-linearity in  Eq.~\eqref{ODE_R}, the first and only pole is simple in the approximation that the two-loop beta function $\beta_2$ is much smaller that one-loop $\beta_1$. Turning on the non-linearity this simple pole induces recursively infinite simple poles (spaced as $\propto 1/\beta_1$) together with logarithm branch-cuts. Ignoring these branch-cuts yields a simple estimate of the renormalons contribution. One starts with
\begin{equation}\label{Y0_Ren}
Y_0(z)=\sum _{i=1}^{\infty } \frac{(-1)^i}{2 i/\beta_1-z} + (\text{analytic terms})\,,
\end{equation}
and by the direct application of the isomorphism in the previous paragraph only $Y_1$ results non-zero. So the Borel-Ecalle resummed renormalons can be written as (recall that the variable $x$ is written in terms of the coupling constant $x=1/ g $)
\begin{equation}
y( g )= y_0( g ) +C e^{-\frac{2}{\beta_1 g }} y_1( g ) = y_0( g )+C\frac{e^{-\frac{2}{\beta_1 g }}}{1+e^{-\frac{2}{\beta_1 g }}}  \,,
\end{equation}
where $y_0( g )$ is just the Cauchy principal value of $Y_0(z)$ in Eq.~\eqref{Y0_Ren} and the purely non-perturbative piece is the one in Eq.~\eqref{explicit_R}.

\bibliographystyle{jhep}
\bibliography{biblio}

\end{document}